\documentclass[%
 reprint,
 amsmath,amssymb,
 aps,
]{revtex4-2}

\usepackage[dvipsnames]{xcolor}
\usepackage{graphicx}
\usepackage{dcolumn}
\usepackage{bm}
\usepackage{tikz}
\usepackage{pgfplots}
\usepackage{subfigure}
\usepackage{multirow}
\usepackage{diagbox}
\usepackage{makecell} 
\usepackage{booktabs} 
\usepackage{float}

 \usetikzlibrary{decorations.markings,backgrounds}
\pgfplotsset{compat=1.16}
\usepgfplotslibrary{fillbetween}

\definecolor{babyblue}{RGB}{131,195,221}
\definecolor{lightblue}{RGB}{48,155,200}
\definecolor{justblue}{RGB}{1,102,169}
\definecolor{deepblue}{RGB}{2,52,107}

\definecolor{babyred}{RGB}{254,134,110}
\definecolor{lightred}{RGB}{252,67,61}
\definecolor{justred}{RGB}{201,24,40}
\definecolor{deepred}{RGB}{111,3,25}

\definecolor{green1}{RGB}{50,146,125}
\definecolor{green2}{RGB}{62,180,155}
\definecolor{green3}{RGB}{112,206,186}

\definecolor{purple3}{RGB}{176,181,219}

\usepackage[colorlinks,
linkcolor=red,
anchorcolor=blue,
citecolor=green
]{hyperref}

\begin{document}

\preprint{APS/123-QED}

\title{Recovery of a generic local Hamiltonian from a degenerate
steady state}

\author{Jing Zhou}

\affiliation{Institute of Physics, Beijing National Laboratory
for Condensed
Matter Physics,\\Chinese Academy of Sciences, Beijing 100190,
China}
	
\affiliation{School of Physical Sciences, University of Chinese
Academy of
  Sciences, Beijing 100049, China}

\author{D. L. Zhou} \email[]{zhoudl72@iphy.ac.cn}
	
\affiliation{Institute of Physics, Beijing National Laboratory
for Condensed
Matter Physics,\\Chinese Academy of Sciences, Beijing 100190,
China}
	
\affiliation{School of Physical Sciences, University of Chinese
Academy of
  Sciences, Beijing 100049, China}

\date{\today}
\begin{abstract}
Hamiltonian Learning (HL) is essential for validating quantum
systems in quantum computing. Not all Hamiltonians can be
uniquely recovered from a steady state. HL success depends
on the Hamiltonian model and steady state. Here, we analyze HL
for a specific type of steady state composed of eigenstates with
degenerate mixing weight, making these Hamiltonian's
eigenstates indistinguishable. To overcome this challenge, we
utilize the orthogonality relationship between the eigenstate
space and its complement space, constructing the orthogonal
space equation. By counting the number of linearly independent
equations derived from a steady state, we determine the
recoverability of a generic local Hamiltonian. Our scheme is
applicable for generic local Hamiltonians under various steady
state, therefore offering a way of measuring the degree to which
a steady state characterizes a Hamiltonian.
\end{abstract}
\maketitle

\section{introduction}
Learning an unknown Hamiltonian of a system from its steady
states or dynamics is a fundamental problem in quantum physics.
Hamiltonian Learning (HL) has numerous applications in quantum
many-body physics~\cite{PhysRevB.99.235109, huang2020predicting,
anshu2021sample, shabani2011estimation,
PhysRevA.79.020305,PhysRevLett.127.200503,9317911}, quantum
state
tomography~\cite{kokail2021entanglement,PhysRevB.107.L100303,PhysRevB.106.L041110,PhysRevA.103.052403,PhysRevC.108.024313}
and quantum device
engineering~\cite{PhysRevA.101.062305,wang2017experimental,
motta2020determining, wiebe2014hamiltonian}. In quantum
many-body physics, the unknown Hamiltonian contains the
information about the interactions in physical system of
interest. Thus it helps us to have a better understanding of the
properties of quantum materials. In quantum state tomography,
the unknown reduced density matrix can be parameterized in terms
of entanglement Hamiltonian with few-body local interactions.
This ansatz is then fitted to a few measurements
~\cite{cotler2020quantum,PhysRevLett.122.150606,dalmonte2018quantum}.
In quantum device engineering, reconstructing the Hamiltonian of
a certain system is crucial for functional testing of the
quantum device~\cite{PRXQuantum.2.010102,PRXQuantum.3.030324}.
The goal of HL is to recover the Hamiltonian with low error
while reducing the number of necessary measurements and
computation resources.

In recent years, many methods of HL have been proposed. These
methods can be divided into two categories based on the
measurements of steady
states~\cite{Qi2019determininglocal,Bairey_2020,PhysRevX.8.021026,PhysRevA.86.022339,Hou_2020,Cao_2020,PhysRevB.100.134201}
or the dynamics of the
systems~\cite{PhysRevA.103.042429,LEE2002349,dutt2021active,li2020hamiltonian,Rattacaso2023highaccuracy,huang2022learning}.
For the steady state, the correlation matrix about the
eigenstates of an unknown Hamiltonian can be utilized to obtain
a candidate Hamiltonian. It is important to note that
entanglement and the exponential growth of the Hilbert space
distinguish quantum many-body systems from classical
systems~\cite{Wang_2015,RevModPhys.81.865}. When an
one-dimensional system is too small, its steady states may not
contain the necessary information to fully recover the
corresponding Hamiltonian. On the other hand, more information
is contained in the dynamical process of many-body Hamiltonians,
which offers higher resolution for the HL\@. However, as time
evolves, the complexity of dynamics grows exponentially,
presenting a challenge to recover Hamiltonian that fits the
measurement result with low computational resources. Introducing
Machine Learning (ML) into HL is an efficient way to solve this
challenge~\cite{dutt2021active}. The essence of HL is querying
the quantum system to generate training data, then interactively
recovering the unknown Hamiltonians. ML is applicable for
constructing query space and extracting information from
training data.

In this work, we focus on HL with measurements on steady states
that are mixed by unknown Hamiltonian's eigenstates with
degeneracy mixing weight. The key idea is to utilize the
orthogonality relationship between the eigenstate space and its
complement space to count the number of LIE obtained from a
certain steady state. We employ the OSE method to analyze the
conditions under which the Hamiltonian can be uniquely
recovered. We then perform numerical computations to verify our
analysis. We address that the ability to recover the Hamiltonian
from a generic steady state depends on the Hamiltonian model and
the degeneracy of the state.

\section{Background}

HL is the process of recovering an unknown Hamiltonian by
monitoring its steady state or dynamics. Assuming the
Hamiltonian $H$ to be recovered is $k$-local, which can be
expanded in terms of a basis of $k$-local Hermitian operators
$\{h_{n}\}$, i.e.,
\begin{equation}
\label{eq:1}
H = \sum_{n=1}^{N}a_{n}h_{n},
\end{equation}
where $h_{n}$ are products of Pauli matrices acting on $k$
contiguous sites. The parameter vector $\vec{a}=[a_{1},
a_{2},\ldots, a_{N}]^{T}$ denotes the non-zero coefficients of
the corresponding terms. The goal of Hamiltonian learning is to
recover the unknown parameter vector $\vec{a}$, and hence the
whole unknown Hamiltonian.

In Ref.~\cite{PhysRevLett.122.020504}, the authors developed
homogeneous operator equations (HOE) method for recovering the
local Hamiltonian from local measurements of a steady state. In
our previous work~\cite{PhysRevA.105.012615}, we applied the HOE
method to recover generic local Hamiltonians with two and three
local interactions, respectively. We found that the
corresponding Hamiltonian can be successfully recovered only
when the chain length $L$ is no less than a critical chain
length $L_{c}$. The critical chain length $L_{c}$ depends on the
Hamiltonian model and the rank $q$ of the steady state. To
determine the critical length $L_{c}$, we introduced the energy
eigenvalue equations (EEE) method to reproduce the results
obtained using the HOE method. By proving the equivalence
between the HOE method and the EEE method, we were able to
analytically determine the critical chain length $L_{c}$ with
the EEE method.

The Hamiltonian we investigated is the two-local Hamiltonian
\begin{equation}
H_2=\sum_{l=1}^L \sum_\eta a_{l \eta}
\sigma_l^\eta+\sum_{l=1}^{L-1} \sum_\eta \sum_\theta a_{l \eta
\theta} \sigma_l^\eta \sigma_{l+1}^\theta,
\end{equation}
and the three-local Hamiltonian 
\begin{equation}
H_3=H_2+\sum_{l=1}^{L-2} \sum_\eta \sum_\theta \sum_\delta a_{l
\eta \theta \delta} \sigma_l^\eta \sigma_{l+1}^\theta
\sigma_{l+2}^\delta.
\end{equation}
The steady state we prepared to be measured is a mixture of $q$
eigenstates of the Hamiltonian to be recovered, which can be
written as
\begin{equation}
\label{eq:4}
\rho =
\sum_{\mu=1}^{q}p_{\mu}|\lambda_{\mu}\rangle\langle\lambda_{\mu}|.
\end{equation}
Here, the mixing weight $p_{\mu}$ is random, and we suppose that
$p_{\alpha}\neq p_{\beta}$ for
$\alpha\neq\beta$. By performing eigen-decomposition on $\rho$,
we can obtain the correct eigenstate $|\lambda_{\mu}\rangle$ of
Hamiltonian to be recovered. Such that we can utilize EEE mothod
to count the number of LIE and determine the critical chain
length $L_{c}$. The critical length $L_{c}$ for the Hamiltonian
model $H_{2}$ and $H_{3}$ under the steady state with rank $q$
is determined by the inequality
\begin{equation}
L_c\left(H_2, \rho\right)=\min _L 2^{L+1} q-q^2-q \geq 12 L-10,
\end{equation}
 \begin{equation}
L_c\left(H_3, \rho\right)=\min _L 2^{L+1} q-q^2-q \geq 39 L-64. \end{equation}
$L_{c}$ is the minimum of $L$ when the inequality holds.

However, our approach to determine the critical chain length
$L_{c}$ is incomplete, it can only be applied to the steady
states without degenerate eigenstates. When there exists
degeneracy in the mixing weight $p_{\mu}$, the eigenstate
becomes indistinguishable, we can only obtain a superposition of
the degenerate eigenstates by eigen-decomposition. In such
cases, the EEE method doesn't work anymore, making the
Hamiltonian recovery more challenging. In this article, we will
address this limitation and present a discussion on how to
determine the critical chain length in cases where the steady
state exhibits degenerate eigenstates. Thus filling the gaps in
the existing researches.

\section{Recovering generic Hamiltonian from a degenerate steady
state}

A state $\rho$ is considered to be a steady state if it's
stationary under the evolution governed by the Hamiltonian $H$.
A steady state can be constructed by mixing Hamiltonian's
eigenstates, as shown in Eq.~\eqref{eq:4}. We refer to $\rho$ as
a degenerate steady state when there exists two or more
eigenstates with the same mixing weight, such that $p_{\alpha} =
p_{\beta}$ for
$\alpha\neq\beta$. In this section, we will discuss recovering
Hamiltonian based on a degenerate steady state.

\subsection{Recovering Hamiltonian from a single-eigenvalued
steady state}

In this subsection, we consider the problem of recovering the
Hamiltonian $H$ from a steady state that is equally mixed by $q$
eigenstates of $H$. We denote such state as single-eigenvalued
steady state, which can be written as
\begin{equation}
\rho_{se} = \sum_{\mu=1}^{q}\frac{1}{q}|E_{\mu}\rangle\langle
E_{\mu}|,
\end{equation}
where $|E_{\mu}\rangle$ is an eigenstate of  $H$.

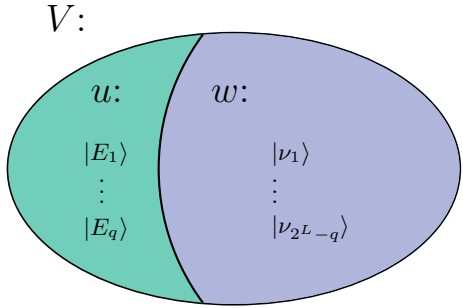
\begin{figure}[h]
\centering
\begin{tikzpicture}
    \coordinate (O) at (0,0);
    \coordinate (B1) at (1,0);
    \coordinate (B2) at (2,0);
    \coordinate (B3) at (3,0);
    \draw[very thick] (O) circle (3cm and 1.8cm);

    \fill[fill=green3] (O) circle (3cm and 1.8cm);
    \begin{scope}
        \path[clip] (B2) circle (3cm);
        \fill[fill=purple3] (O) circle (3cm and 1.8cm);
    \end{scope}
    \begin{scope}
        \clip (O) circle (3cm and 1.8cm);
        \draw[thick] (B2) circle (3cm and 3cm);
    \end{scope}
\node[text width=3cm] at (-1,2) 
    { \Large{$V$: }};
\node[text width=3cm] at (-0.4,1) 
    { \Large{$u$: }};
\node[text width=3cm] at (-0.5,0.2) 
    { $|E_{1}\rangle$};
\node[text width=3cm] at (-0.3,-0.2) 
    { $\vdots$};
\node[text width=3cm] at (-0.5,-0.8) 
    { $|E_{q}\rangle$};
\node[text width=3cm] at (1.2,1) 
    { \Large{$w$: }};
\node[text width=3cm] at (2,0.2) 
    { $|\nu_{1}\rangle$};
\node[text width=3cm] at (2,-0.2) 
    { $\vdots$};
\node[text width=3cm] at (2,-0.8) 
    { $|\nu_{2^{L}-q}\rangle$};
\end{tikzpicture}
\caption{The relationship between space $V$, $u$ and
$w$.}\label{fig:1}
\end{figure}

We denote the Hilbert space spanned by all eigenstates of $H$ as
$V$, and its subspace spanned by
$\{|E_{\mu}\rangle\}_{\mu=1}^{q}$ as $u$. Noting that $u$ is an
invariant subspace of $H$.
The eigen-decomposition of $\rho$ gives $q$ eigenvectors
$\{|\lambda_{m}\rangle\}_{m=1}^{q}$ with a single eigenvalue
$1/q$. $|\lambda_{m}\rangle$ is contained in the space $u$,
therefore it can be expressed as a linear combination of
$\{|E_{n}\rangle\}_{n=1}^{q}$,
\begin{equation}
|\lambda_{m}\rangle = \sum_{n=1}^{q}c_{n}|E_{n}\rangle,
\end{equation}
Next we consider the complementary space of $u$ with respect to
the space $V$, which is denoted as $w$, $w=V-u$. We constitute
the orthogonal basis $\{\nu_{l}\}_{l=1}^{2^{L}-q}$ of $w$. The
space $u$ and $w$ are orthogonal to each other,
\begin{equation}
\forall l, m: \langle\nu_{l}|\lambda_{m}\rangle = 0.
\end{equation}
The relationship between space $V$, $u$ and $w$ is shown in
Fig.~\ref{fig:1}.

As mentioned above, $u$ is an invariant subspace of $H$.
Therefore acting $H$ on any vector in $u$ will result in another
vector $H|\lambda_{m}\rangle$ that is also contained within the
space $u$. This property allows us to write a set of homogeneous
linear equations for the unknowns $\vec{a}$,
\begin{equation}
\label{eq:10}
\forall\ l, m:
\sum_{n}a_{n}\langle\nu_{l}|h_{n}|\lambda_{m}\rangle=0.
\end{equation}
From Eqs.~\eqref{eq:10}, we can derive a set of $S =
2q\cdot(2^{L}-q)$ real-valued LIE for $\vec{a}$. Since
Hamiltonian model described by Eq.~\eqref{eq:1} has $N$ unknown
parameters, it can possibly be recovered from the steady state
$\rho_{se}$ in the form of $\alpha\vec{a}$ when $S\geq N-1$,
with $\alpha $ being any real number. We refer to this method as
the orthogonal space equations (OSE) method.

The procedure for recovering the Hamiltonian using the OSE
method consists of the following steps. STEP 1: We generate a
series of random Hamiltonians $H_{2}$ and $H_{3}$ by sampling
the coefficient $a_{i} \sim N(0,1)$ from a Gaussian
distribution. STEP 2: A steady state is prepared by mixing
random Hamiltonian's eigenstates, where the mixing weight is
specified. STEP3: Performing eigen-decomposition on the steady
state from Step 2, we obtain a set of vectors that share the
same eigenvalue. These vectors collectively form the space $u$.
STEP 4: Constructing orthogonal vectors in space $w$ that are
linearly independent of the vectors in the complement space $u$.
STEP 5: Building the homogeneous linear equations given by
Eqs.~\eqref{eq:10} from the vectors in space $u$ and $w$. Then
we use the numpy function np.linalg.svd to solve the equations
and obtain the recovered coefficient $\vec{a}$.

\begin{figure}[h]
\subfigbottomskip=10pt
\subfigure[]{
	\begin{tikzpicture}[align=center]
	\begin{semilogyaxis}
	[
	small, width=6.8cm, 
	xlabel = $L$,
	ylabel=$\Delta$,
	ytick style = {right},
	ytick={0,1e-17,1e-15,1e-13,1e-11,1e-9,1e-7,1e-5,1},
	xtick={1,2,3,4,5,6,7,8,9, 10},
	legend style = {at={(0.88,0.9)}},
legend entries = {[font=\tiny]$H_{2}$ q=2,[font=\tiny]$H_{2}$
q=3}]
	
	\addplot[mark=square*,deepblue,
	error bars/.cd, 
	x dir=both, x explicit,
	y dir=both, y explicit,
	] coordinates {

(2,0.805883944890049)+=(0,0.3905755752964992)-=(0,0.412964818694321)
(3,0.8269952095006995)+=(0,0.27521562012377465)-=(0,0.3624300246085877)
(4,2.1865738492339854e-14)+=(0,1.196542416747998e-13)-=(0,1.8941339413273397e-14)
(5,3.527923392159418e-14)+=(0,5.619455312975188e-13)-=(0,3.0745626691142534e-14)(6,4.0715574376045555e-14)+=(0,2.197707997974794e-13)-=(0,3.497914802890851e-14)(7,9.391986248365845e-14)+=(0,1.857269397403603e-12)-=(0,8.103614234330178e-14)
(8,1.981032997861651e-13)+=(0,4.859581413690393e-12)-=(0,1.8287783276119527e-13)(9,5.153334503424568e-13)+=(0,9.76236448886667e-12)-=(0,4.882330463060147e-13)
(10,2.2968026897332528e-12)+=(0,8.270203347912772e-12)-=(0,2.24732435583587e-12)
};
	\addplot[mark=*,babyblue,
	error bars/.cd, 
	x dir=both, x explicit,
	y dir=both, y explicit,
	] coordinates {

(2,0.8412180362305941)+=(0,0.32748114833038455)-=(0,0.3814109531348314)
(3,1.1818916611054939e-14)+=(0,4.8392651167427705e-14)-=(0,7.910784429820938e-15)
(4,8.459654494489881e-15)+=(0,4.4216967125565745e-14)-=(0,5.928232960606554e-15)(5,1.2241936515840624e-14)+=(0,9.7054347270179e-14)-=(0,9.082705004613181e-15)
(6,2.2198312644831833e-14)+=(0,2.5588229889217113e-13)-=(0,1.8697675377846587e-14)
(7,3.529992770147745e-14)+=(0,3.3088157671731647e-13)-=(0,3.0316229239582637e-14)
(8,5.213643899521525e-14)+=(0,3.7018616836998813e-13)-=(0,4.297119528727604e-14)(9,1.5622076623142245e-13)+=(0,1.0561908002566302e-12)-=(0,1.4365871380936169e-13)
(10,1.37185753376702e-12)+=(0,7.631958844790087e-12)-=(0,1.342899800546809e-12)
	};
		
	\end{semilogyaxis}
	\end{tikzpicture}
	}

\subfigure[]{
	\begin{tikzpicture}[align=center]
	\begin{semilogyaxis}
	[
	small, width=6.8cm, 
	xlabel = $L$,
	ylabel=$\Delta$,
	ytick style = {right},
	ytick={0,1e-17,1e-15,1e-13,1e-11,1e-9,1e-7,1e-5,1},
	xtick={1,2,3,4,5,6,7,8,9,10},
	legend style = {at={(0.88,0.9)}},
legend entries = {[font=\tiny]$H_{3}$ q=2,[font=\tiny]$H_{3}$
q=3}
	]	
	\addplot[mark=square*,deepred,
	error bars/.cd, 
	x dir=both, x explicit,
	y dir=both, y explicit,
	] coordinates {
	
(3,0.8393483104076568)+=(0,0.12102334602201881)-=(0,0.22244426698210584)
(4,0.8497465832900636)+=(0,0.08179558136626985)-=(0,0.14614839721426098)
(5,0.849215076516269)+=(0,0.06591134944506671)-=(0,0.11405966809683599)
(6,6.664113204088316e-14)+=(0,1.0400948791401553e-13)-=(0,4.385608847963012e-14)(7,6.835643633132757e-14)+=(0,1.7998841018554942e-13)-=(0,4.685116068444501e-14)(8,9.175131431173856e-14)+=(0,2.441069812165995e-13)-=(0,6.42999477727771e-14)
(9,3.0023291022354093e-13)+=(0,3.472859266272289e-12)-=(0,2.4980953835521606e-13)
(10,2.1133221517171407e-12)+=(0,5.740077790433959e-12)-=(0,2.0172967406020283e-12)
	};
	\addplot[mark=*, babyred,
	error bars/.cd, 
	x dir=both, x explicit,
	y dir=both, y explicit,
	] coordinates {
(3,0.8352213267857806)+=(0,0.11652390887120845)-=(0,0.1528075305936616)
(4,0.8510227472658525)+=(0,0.08913180418835098)-=(0,0.10554688444294025)
(5,4.126955079568714e-14)+=(0,6.474584003341276e-14)-=(0,2.8210424382207823e-14)(6,3.019302135811691e-14)+=(0,6.145691926966856e-14)-=(0,1.8800233724598196e-14)(7,3.973668286585143e-14)+=(0,8.808963654645379e-14)-=(0,2.7580304958632118e-14)(8,5.3643081005087496e-14)+=(0,9.854337044620882e-14)-=(0,3.676060293362686e-14)(9,1.3289674461253166e-13)+=(0,1.5902354258160998e-12)-=(0,1.0991845027155724e-13)
(10,1.4506521385223596e-12)+=(0,5.702079007014763e-12)-=(0,1.3993039188906992e-12)
	};	
	\end{semilogyaxis}
	\end{tikzpicture}
}\\
\caption{ We reconstruct (a) $H_{2}$ and (b) $H_{3}$ by means of
OSE from a single eigen-valued degenerate steady state.
Simulations are executed over $200$ random Hamiltonians with two
different states for
each system size $L$. The squares and circles represent states
of q=2 and 3, respectively.
}\label{fig:2}
\end{figure}
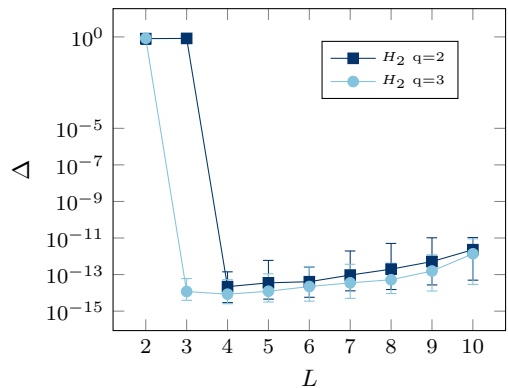
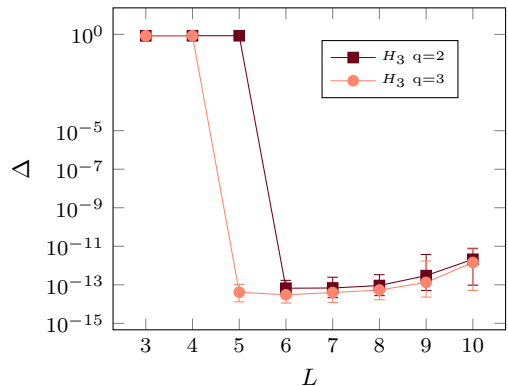

Following the analysis above, we numerically obtain
reconstructing errors of Hamiltonian $H_{2}$ and $H_{3}$ from
the steady state $\rho_{se}$ with chain length $L$ ranging from
2 to 10, as shown in Fig.~\ref{fig:2}. We find that the OSE
method successfully recovers the Hamiltonian when $L$ is equal
to or larger than a critical value $L_{c}$, which depends on the
Hamiltonian model and rank of the steady state. The value of
$L_{c}$ can be determined by
\begin{equation}
L_{c}(H,\rho_{se})=\min _L2q\cdot(2^{L}-q)\geq N-1,
\end{equation}
where $q$ is the rank of degenerate steady state $\rho_{se}$,
$N$ is the number of terms of generic local Hamiltonian $H$.

It can be inferred that when $q$ is larger, more information is
contained in the steady state, therefore more Hamiltonian can be
uniquely recovered. Due to having more unknowns than $H_{2}$ at
the same chain length, recovering $H_{3}$ from measurements on a
steady state becomes more challenging. Comparing with Fig.~1 in
Ref.~\cite{PhysRevA.105.012615}, we find that it is harder to
recover the Hamiltonian with a degenerate steady state. This
difficulty arises from the inability to effectively extract the
unknown Hamiltonian's eigenstates from the degenerate steady
state. Instead of obtaining individual eigenstates, we can only
acquire a superposition of these eigenstates. Consequently, in
the process of eigen-decomposition of degenerate steady state,
the information regarding specific eigenstates is lost, making
it harder to recover the Hamiltonian.

\subsection{Recovering Hamiltonian from a multi-eigenvalued
steady state}
In this subsection, we consider recovering Hamiltonian $H$ from
a multi-eigenvalued steady state.

\begin{figure}[h]
\centering
\begin{tikzpicture}
    \coordinate (O) at (0,0);
    \coordinate (B1) at (1.2,0);
    \coordinate (B2) at (2.4,0);
    \coordinate (B3) at (3.6,0);
    \draw[very thick] (O) circle (3cm and 1.8cm);

    \fill[fill=green1] (O) circle (3cm and 1.8cm);

    \begin{scope}
        \path[clip] (B1) circle (3cm);
        \fill[fill=green2] (O) circle (3cm and 1.8cm);
    \end{scope}
    \begin{scope}
        \path[clip] (B2) circle (3cm);
        \fill[fill=green3] (O) circle (3cm and 1.8cm);
    \end{scope}
    \begin{scope}
        \path[clip] (B3) circle (3cm);
        \fill[purple3] (O) circle (3cm and 1.8cm);
    \end{scope}
    \begin{scope}
        \clip (O) circle (3cm and 1.8cm);
        \draw[thick] (B1) circle (3cm and 3cm);
        \draw[thick] (B2) circle (3cm and 3cm);
        \draw[thick] (B3) circle (3cm and 3cm);
    \end{scope}
\node[text width=3cm] at (-1,2) 
    { \Large{$V$: }};
\node[text width=3cm] at (-1,0.8) 
    { \large{$u_{1}$: }};
\node[text width=3cm] at (-1.2,0.3) 
    { $|E_{1}^{1}\rangle$ };
\node[text width=3cm] at (-1,0) 
    { $\vdots$ };
\node[text width=3cm] at (-1.2,-0.6) 
    { $|E_{1}^{q_{1}}\rangle$ };
\node[text width=3cm] at (-0.2, 0) 
    { $\cdots$ };

\node[text width=3cm] at (1.2,0.8) 
    { \large{$u_{M}$: }};
\node[text width=3cm] at (1.1,0.3) 
    { $|E_{M}^{1}\rangle$ };
\node[text width=3cm] at (1.3,0) 
    { $\vdots$ };
\node[text width=3cm] at (1.1,-0.6) 
    { $|E_{M}^{q_{M}}\rangle$ };

\node[text width=3cm] at (2.4,0.8) 
    { \large{$W$: }};
\node[text width=3cm] at (2.9,0.3) 
    { $|\nu_{1}\rangle$ };
\node[text width=3cm] at (3.1,0) 
    { $\vdots$ };
\node[text width=3cm] at (2.9,-0.6) 
    { $|\nu_{2^{L}-Q}\rangle$ };
\end{tikzpicture}
\caption{The relationship between space $V$, $U$ and $W$}
    \label{fig:3}
\end{figure}
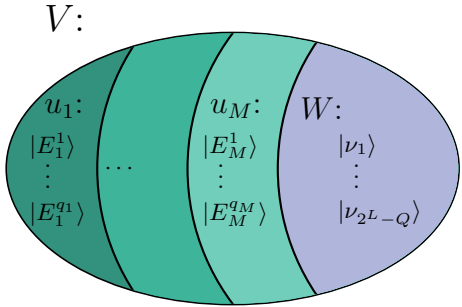

Suppose that the steady state $\rho_{me}$ has eigenvalues
$\{p_{m}\}_{m=1}^{M}$ with corresponding degeneracy
$\{q_{m}\}_{m=1}^{M}$, we can express $\rho_{me}$ as
\begin{equation}
\rho_{me} =
\sum_{m=1}^{M}p_{m}\sum_{j=1}^{q_{m}}|E_{m}^{j}\rangle\langle
E_{m}^{j}|,
\end{equation}
where $|E_{m}^{j}\rangle$ is an eigenstate of $H$. We denote the
Hilbert space spanned by all eigenstates of $H$ as $V$, and its
subspace spanned by eigenstates with the same weight $p_{m}$,
i.e., $\{|E_{m}^{j}\rangle\}_{j=1}^{q}$ as $u_{m}$. Sum of the
$u_{m}$ forms the set $U$, each element of $U$ is an invariant
subspace of $H$. We refer to the complement space of $U$ with
respect to $V$ as $W$, $W=V-U$.
The eigen-decomposition of $\rho_{me}$ gives $M$ eigenvalues
$\{p_{m}\}_{m=1}^{M}$, each with corresponding $q_{m}$
eigenstates $\{|\lambda_{m}^{j}\rangle\}_{j=1}^{q_{m}}$.
$|\lambda_{m}^{j}\rangle$ is contained in the space $u_{m}$,
which means it can be expressed as the linear combination of
vectors in $\{|E_{m}^{j}\rangle\}_{j=1}^{q_{m}}$. The
relationship between these spaces is shown in Fig.~\ref{fig:3}.

To obtain more information about $H$, we construct an orthogonal
basis $\{\nu_{l}\}_{l=1}^{2^{L}-Q}$ of the space $W$, where $Q$
is the number of eigenstate of $\rho_{me}$, given by $Q =
\sum_{m}q_{m}$. Acting $H$ on $|\lambda_{m}^{j}\rangle$, the
resulting vector $H|\lambda_{m}^{j}\rangle$ lies in the
invariant subspace $u_{m}$, and vectors in different subspace
are orthogonal to each other. The orthogonal relationships above
lead to following set of homogeneous linear equations for
$\vec{a}$,
\begin{subequations}
\label{eq:12}
\begin{equation}
\label{eq:12a}
\forall j, m\neq m^{\prime}:
\sum_{n}a_{n}\langle\lambda_{m^{\prime}}^{i}|h_{n}|\lambda_{m}^{j}\rangle=0,
\end{equation}
\begin{equation}
\label{eq:12b}
\forall j, m, l:
\sum_{n}a_{n}\langle\nu_{l}|h_{n}|\lambda_{m}^{j}\rangle=0.
\end{equation}
\end{subequations}
Combining Eqs.~\eqref{eq:12}, we can get 
\begin{equation}
\label{eq:13}
S = 2\cdot\sum_{m=1}^{M}q_{m}\sum_{m^{\prime}\geq
m}^{M}q_{m^{\prime}}+2Q(2^{L}-Q)
\end{equation}
real-valued equations. Such that $S$ is the number of LIE one
obtains from a degenerate steady state. When $S\geq N-1$, the
number of LIE surpasses the degree of freedom of the unknowns,
consequently we can successfully recover the Hamiltonian from
the steady state.

\begin{figure}[h]
\subfigbottomskip=10pt
\subfigure[]{
	\begin{tikzpicture}[align=center]
	\begin{semilogyaxis}
	[
	small, width=6.8cm, 
	xlabel = $L$,
	ylabel=$\Delta$,
	ytick style = {right},
	ytick={0,1e-17,1e-15,1e-13,1e-11,1e-9,1e-7,1e-5,1},
	xtick={1,2,3,4,5,6,7,8,9, 10},
	legend style = {at={(0.88,0.9)}},
	legend entries = {[font=\tiny]$H_{2}$, [font=\tiny]$H_{3}$}
	]
	
	\addplot[mark=square*, justblue,
	error bars/.cd, 
	x dir=both, x explicit,
	y dir=both, y explicit,
	] coordinates {
(2,0.7959326445502953)+=(0,0.2348207069061311)-=(0,0.4000656360509436)
(3,5.24075944156929e-15)+=(0,4.457306359537594e-15)-=(0,2.960521024727335e-15)
(4,5.136177326839529e-15)+=(0,7.035057007365865e-15)-=(0,3.13763885610415e-15)
(5,6.803387655806722e-15)+=(0,2.135740834455376e-14)-=(0,5.0302364993490494e-15)(6,1.0197352448519618e-14)+=(0,2.1054151805071978e-14)-=(0,7.35454603578856e-15)(7,1.673803125821752e-14)+=(0,5.219314464106213e-14)-=(0,1.261888683105514e-14)
(8,2.6508755501869546e-14)+=(0,9.463882128309868e-14)-=(0,2.0493848503131517e-14)
(9,5.962393127775907e-14)+=(0,4.734701674571498e-13)-=(0,4.68947020176656e-14)
(10,7.943866622068552e-13)+=(0,3.937788260660232e-12)-=(0,7.746919960377045e-13)
};
	
	\addplot[mark=*, justred,
	error bars/.cd, 
	x dir=both, x explicit,
	y dir=both, y explicit,
	] coordinates {	
(3,0.8368254990715583)+=(0,0.11214657676936701)-=(0,0.1558953089374956)
(4,0.8507985618884917)+=(0,0.07901557937327142)-=(0,0.10463465065248756)
(5,1.749331968916946e-14)+=(0,2.5696257999911255e-14)-=(0,1.164208614542525e-14)(6,1.792407317404432e-14)+=(0,3.147913584086487e-14)-=(0,1.112628985309419e-14)
(7,2.4901505111398737e-14)+=(0,5.368473295896544e-14)-=(0,1.7010633938428972e-14)
(8,4.304220788418069e-14)+=(0,1.1281480741681378e-13)-=(0,2.902450323558347e-14)(9,8.037101806872101e-14)+=(0,5.509273462520746e-13)-=(0,6.248943856650407e-14)
(10,9.129988735621939e-13)+=(0,2.7573555863169445e-12)-=(0,8.767241738507977e-13)
	};
	\end{semilogyaxis}
	\end{tikzpicture}
}\\
\caption{ We reconstruct $H_{2}$ and $H_{3}$ by means of OSE
from $\rho_{me}$.
Simulations are executed over $200$ random Hamiltonians for each
system size $L$. }\label{fig:4}
\end{figure}
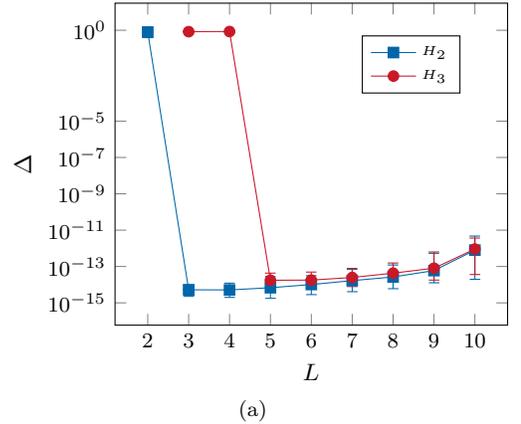

Next, we numerically recover the generic local Hamiltonian
$H_{2}$ and $H_{3}$ with chain length $L$ ranging from 2 to 10
based on multiple-eigenvalued steady state,
\begin{equation}
\rho_{me}=0.2(|\psi_{0}\rangle\langle\psi_{0}| +
|\psi_{1}\rangle\langle\psi_{1}| ) +
0.3(|\psi_{2}\rangle\langle\psi_{2}| +
|\psi_{3}\rangle\langle\psi_{3}| ),
\end{equation}
where $|\psi_{n}\rangle$ is the $n$-th eigenstate of the
Hamiltonian to be recovered. Fig.~\ref{fig:4} presents the
reconstructing error as a function of chain length for
Hamiltonian $H_{2}$ and $H_{3}$. The critical chain length of
$H_{2}$ and $H_{3}$ under the state $\rho_{me}$ is 3 and 5,
respectively. Eigen-decomposing state $\rho_{me}$, one obtains
two eigenvalues along with their corresponding two eigenstates,
resulting in a mixing weight degeneracy $\textbf{q}=(2,2)$. To
determine the number of LIE obtained from $\rho_{me}$, we
utilize Eq.~\eqref{eq:13}, yielding $S = 2^{L+3}-24$. Numerical
results align with the criteria that when $S\geq N-1$,
Hamiltonian $H_{2}$ and $H_{3}$ can be successfully recovered.

To further verify the validity of Eq.~\eqref{eq:13} in
determining the number of LIE obtained from a degenerate steady
state, we compare the value of $S$ with $r =\text{rank}G$
obtained through the HOE method. We consider the
multi-eigenvalued steady state $\rho_{me}$ of the Hamiltonians
$H_{2}$ and $H_{3}$. Table.~\ref{table:1} presents the value of
$N$, $S$ and $r$ as a function of $L$ for Hamiltonian $H_{2}$
and $H_{3}$. When the number of LIE given by Eq.\eqref{eq:13} is
lower than the number of Hamiltonian terms, i.e., $S<N$, $r$ is
equal to $S$ and the Hamiltonian cannot be uniquely recovered.
Under such condition, the consistency between the values of $S$
and $r$ confirms the validity of Eqs.\eqref{eq:13}. As the chain
length $L$ increases, $S$ grows exponentially, while $N$
increases polynomially. Once $S$ exceeds $N$, $r$ becomes equal
to $N-1$, enabling the successful recovery of Hamiltonians
$H_{2}$ and $H_{3}$.

Finally, we conclude that, when the multi-eigenvalued steady
state $\rho_{me}$ satisfies the inequality below, it contains
the information required to recover the generic local
Hamiltonian $H_{2}$ or $H_{3}$,
\begin{equation}
2\cdot\sum_{m=1}^{M}q_{m}\sum_{m^{\prime}\geq
m}^{M}q_{m^{\prime}}+2Q(2^{L}-Q) \geq N-1,
\end{equation}
where $N=12L-9$ and $48L-81$ for $H_{2}$ and $H_{3}$,
respectively.

\begin{table}[htbp]
\renewcommand{\arraystretch}{1.3}\centering
\setlength{\tabcolsep}{0mm}{
  \setlength{\tabcolsep}{3mm}
    \begin{tabular}{cccccc} \toprule
\quad & \quad& \multicolumn{2}{c}{$H_{2}$}&
\multicolumn{2}{c}{$H_{3}$}\\
     \cmidrule(r){3-4}\cmidrule(lr){5-6}

      \ \ $L$\ \ & \ \ S\ \ &N&r&N&r\\

      2&8&15&8&/&/\\

      3&40&27&26&63&40\\
      
      4&104&39&38&111&104\\
      
      5&232&51&50&159&158\\
      
      6&488&63&62&207&206\\
      
      7&1000&75&74&255&254\\
      
      8&2024&87&86&303&302\\
      
      9&4072&99&98&351&350\\
      
      10&8186&111&110&399&398\\
\toprule

    \end{tabular}}
\caption{$S$, $N$ and $r$ as the function of $L$ and $H$.
}\label{table:1}
\end{table}

\section{summary and outlook}
In this paper, we have presented a novel scheme for evaluating
the degree to which a degenerate steady state characterizes a
Hamiltonian. Our approach leverages the use of OSE to quantify
the number of LIE obtained from a degenerate steady state,
therefore offering valuable insight into the properties of the
underlying Hamiltonian. As a proof of principle, we applied this
method to both single and multiple-eigenvalued steady state of
generic local Hamiltonians. The numerical results showed that
the Hamiltonian can only be recovered when the chain length $L$
is greater than or equal to the critical chain length $L_{c}$,
which is determined by the Hamiltonian model and the degeneracy
of the steady state. To verify the correctness on number of LIE
counted from the OSE, we compare the value of $S$ and $r$. The
consistency between these two values suggests the validity of
the OSE method. Finally, we analytically determine the value of
$L_{c}$ from the inequality relations between number of LIE and
the Hamiltonian terms.

A natural extension of our approach for HL is the recovery of
Hamiltonian with symmetries. In contrast to generic local
Hamiltonian, the symmetric Hamiltonian model has less unknowns,
here one deals with degenerate steady states as well as
degenerate Hamiltonian energy levels. It has more elaborate
space structure.

In conclusion, our work presents a scalable recipe for counting
LIE of a specific steady state, providing a promising direction
for further exploration. This research opens up possibilities
for investigating unknown symmetries and advancing our
understanding of many-body Hamiltonians.

\section{acknowledgments}
This work is supported by National Key Research and Development Program of China
(Grant No. 2021YFA0718302 and No. 2021YFA1402104), National Natural Science
Foundation of China (Grants No. 12075310), and the Strategic Priority Research
Program of Chinese Academy of Sciences (Grant No. XDB28000000).

\bibliographystyle{apsrev4-2} \bibliography{Ham_tomo}

\end{document}